\documentclass[aps,prl,superscriptaddress,showpacs,amssymb,amsmath,amsfonts,twocolumn,floatfix]{revtex4-1}

\usepackage{amsmath}
\usepackage{amssymb}
\usepackage{soul}

\usepackage{enumerate}
\usepackage{graphicx}
\usepackage{dcolumn}
\usepackage{bm}
\usepackage{xcolor}

\begin{document} 
\title{Target and Double Spin Asymmetries of Deeply Virtual $\pi^0$ Production with a Longitudinally Polarized Proton Target and CLAS}

 \date{\today}

\newcommand*{\ANL}{Argonne National Laboratory, Argonne, Illinois 60439}
\newcommand*{\ANLindex}{1}
%\affiliation{\ANL}
\newcommand*{\ASU}{Arizona State University, Tempe, Arizona 85287-1504}
\newcommand*{\ASUindex}{2}
%\affiliation{\ASU}
\newcommand*{\CSUDH}{California State University, Dominguez Hills, Carson, CA 90747}
\newcommand*{\CSUDHindex}{3}
%\affiliation{\CSUDH}
\newcommand*{\CMU}{Carnegie Mellon University, Pittsburgh, Pennsylvania 15213}
\newcommand*{\CMUindex}{4}
%\affiliation{\CMU}
\newcommand*{\CUA}{Catholic University of America, Washington, D.C. 20064}
\newcommand*{\CUAindex}{5}
%\affiliation{\CUA}
\newcommand*{\SACLAY}{CEA, Centre de Saclay, Irfu/Service de Physique Nucl\'eaire, 91191 Gif-sur-Yvette, France}
\newcommand*{\SACLAYindex}{6}
%\affiliation{\SACLAY}
\newcommand*{\CNU}{Christopher Newport University, Newport News, Virginia 23606}
\newcommand*{\CNUindex}{7}
%\affiliation{\CNU}
\newcommand*{\UCONN}{University of Connecticut, Storrs, Connecticut 06269}
\newcommand*{\UCONNindex}{8}
%\affiliation{\UCONN}
\newcommand*{\FU}{Fairfield University, Fairfield CT 06824}
\newcommand*{\FUindex}{9}
%\affiliation{\FU}
\newcommand*{\FIU}{Florida International University, Miami, Florida 33199}
\newcommand*{\FIUindex}{10}
%\affiliation{\FIU}
\newcommand*{\FSU}{Florida State University, Tallahassee, Florida 32306}
\newcommand*{\FSUindex}{11}
%\affiliation{\FSU}
\newcommand*{\Genova}{Universit$\grave{a}$ di Genova, 16146 Genova, Italy}
\newcommand*{\Genovaindex}{12}
%\affiliation{\Genova}
\newcommand*{\GWUI}{The George Washington University, Washington, DC 20052}
\newcommand*{\GWUIindex}{13}
%\affiliation{\GWUI}
\newcommand*{\ISU}{Idaho State University, Pocatello, Idaho 83209}
\newcommand*{\ISUindex}{14}
%\affiliation{\ISU}
\newcommand*{\INFNFE}{INFN, Sezione di Ferrara, 44100 Ferrara, Italy}
\newcommand*{\INFNFEindex}{15}
%\affiliation{\INFNFE}
\newcommand*{\INFNFR}{INFN, Laboratori Nazionali di Frascati, 00044 Frascati, Italy}
\newcommand*{\INFNFRindex}{16}
%\affiliation{\INFNFR}
\newcommand*{\INFNGE}{INFN, Sezione di Genova, 16146 Genova, Italy}
\newcommand*{\INFNGEindex}{17}
%\affiliation{\INFNGE}
\newcommand*{\INFNRO}{INFN, Sezione di Roma Tor Vergata, 00133 Rome, Italy}
\newcommand*{\INFNROindex}{18}
%\affiliation{\INFNRO}
\newcommand*{\INFNTUR}{INFN, Sezione di Torino, 10125 Torino, Italy}
\newcommand*{\INFNTURindex}{19}
%\affiliation{\INFNTUR}
\newcommand*{\ORSAY}{Institut de Physique Nucl\'eaire, CNRS/IN2P3 and Universit\'e Paris Sud, Orsay, France}
\newcommand*{\ORSAYindex}{20}
%\affiliation{\ORSAY}
\newcommand*{\ITEP}{Institute of Theoretical and Experimental Physics, Moscow, 117259, Russia}
\newcommand*{\ITEPindex}{21}
%\affiliation{\ITEP}
\newcommand*{\JMU}{James Madison University, Harrisonburg, Virginia 22807}
\newcommand*{\JMUindex}{22}
%\affiliation{\JMU}
\newcommand*{\KNU}{Kyungpook National University, Daegu 702-701, Republic of Korea}
\newcommand*{\KNUindex}{23}
%\affiliation{\KNU}
\newcommand*{\LPSC}{LPSC, Universit\'e Grenoble-Alpes, CNRS/IN2P3, Grenoble, France}
\newcommand*{\LPSCindex}{24}
%\affiliation{\LPSC}
\newcommand*{\MISS}{Mississippi State University, Mississippi State, MS 39762-5167}
\newcommand*{\MISSindex}{25}
%\affiliation{\MISS}
\newcommand*{\UNH}{University of New Hampshire, Durham, New Hampshire 03824-3568}
\newcommand*{\UNHindex}{26}
%\affiliation{\UNH}
\newcommand*{\NSU}{Norfolk State University, Norfolk, Virginia 23504}
\newcommand*{\NSUindex}{27}
%\affiliation{\NSU}
\newcommand*{\OHIOU}{Ohio University, Athens, Ohio  45701}
\newcommand*{\OHIOUindex}{28}
%\affiliation{\OHIOU}
\newcommand*{\ODU}{Old Dominion University, Norfolk, Virginia 23529}
\newcommand*{\ODUindex}{29}
%\affiliation{\ODU}
\newcommand*{\RPI}{Rensselaer Polytechnic Institute, Troy, New York 12180-3590}
\newcommand*{\RPIindex}{30}
%\affiliation{\RPI}
\newcommand*{\URICH}{University of Richmond, Richmond, Virginia 23173}
\newcommand*{\URICHindex}{31}
%\affiliation{\URICH}
\newcommand*{\ROMAII}{Universita' di Roma Tor Vergata, 00133 Rome Italy}
\newcommand*{\ROMAIIindex}{32}
%\affiliation{\ROMAII}
\newcommand*{\MSU}{Skobeltsyn Institute of Nuclear Physics, Lomonosov Moscow State University, 119234 Moscow, Russia}
\newcommand*{\MSUindex}{33}
%\affiliation{\MSU}
\newcommand*{\SCAROLINA}{University of South Carolina, Columbia, South Carolina 29208}
\newcommand*{\SCAROLINAindex}{34}
%\affiliation{\SCAROLINA}
\newcommand*{\TEMPLE}{Temple University,  Philadelphia, PA 19122 }
\newcommand*{\TEMPLEindex}{35}
%\affiliation{\TEMPLE}
\newcommand*{\JLAB}{Thomas Jefferson National Accelerator Facility, Newport News, Virginia 23606}
\newcommand*{\JLABindex}{36}
%\affiliation{\JLAB}
\newcommand*{\UTFSM}{Universidad T\'{e}cnica Federico Santa Mar\'{i}a, Casilla 110-V Valpara\'{i}so, Chile}
\newcommand*{\UTFSMindex}{37}
%\affiliation{\UTFSM}
\newcommand*{\EDINBURGH}{Edinburgh University, Edinburgh EH9 3JZ, United Kingdom}
\newcommand*{\EDINBURGHindex}{38}
%\affiliation{\EDINBURGH}
\newcommand*{\GLASGOW}{University of Glasgow, Glasgow G12 8QQ, United Kingdom}
\newcommand*{\GLASGOWindex}{39}
%\affiliation{\GLASGOW}
\newcommand*{\VT}{Virginia Tech, Blacksburg, Virginia   24061-0435}
\newcommand*{\VTindex}{40}
%\affiliation{\VT}
\newcommand*{\VIRGINIA}{University of Virginia, Charlottesville, Virginia 22901}
\newcommand*{\VIRGINIAindex}{41}
%\affiliation{\VIRGINIA}
\newcommand*{\WM}{College of William and Mary, Williamsburg, Virginia 23187-8795}
\newcommand*{\WMindex}{42}
%\affiliation{\WM}
\newcommand*{\YEREVAN}{Yerevan Physics Institute, 375036 Yerevan, Armenia}
\newcommand*{\YEREVANindex}{43}
%\affiliation{\YEREVAN}

\newcommand*{\NOWCNU}{Christopher Newport University, Newport News, Virginia 23606}
\newcommand*{\NOWUK}{University of Kentucky, LEXINGTON, KENTUCKY 40506}
\newcommand*{\NOWJLAB}{Thomas Jefferson National Accelerator Facility, Newport News, Virginia 23606}
\newcommand*{\NOWZZZ}{unused, unused}
\newcommand*{\NOWODU}{Old Dominion University, Norfolk, Virginia 23529}
\newcommand*{\NOWINFNGE}{INFN, Sezione di Genova, 16146 Genova, Italy}
\newcommand*{\NOWEDINBURGH}{Edinburgh University, Edinburgh EH9 3JZ, United Kingdom}
 %%%%%%%%%%%%%%% END OF Latex Macros for institute addresses  %%%%%%%%%%%%%%%%%%%%%%%%% 

\author{A.~Kim}
\affiliation{\UCONN}
\affiliation{\KNU}
\author {H.~Avakian}
\affiliation{\JLAB}
\author {V.~Burkert}
\affiliation{\JLAB}
\author {K.~Joo}
\affiliation{\UCONN}
\author {W.~Kim} 
\affiliation{\KNU}

\author {K.P. ~Adhikari} 
\affiliation{\MISS}
\affiliation{\ODU}
\author {Z.~Akbar} 
\affiliation{\FSU}
\author {S. ~Anefalos~Pereira} 
\affiliation{\INFNFR}
\author {R.A.~Badui} 
\affiliation{\FIU}
\author {M.~Battaglieri} 
\affiliation{\INFNGE}
% ...

\collaboration{The CLAS Collaboration}
\noaffiliation

\begin{abstract}
The target and double spin asymmetries of the exclusive pseudoscalar channel $\vec e\vec p\to ep\pi^0$ were measured for the first time in the deep-inelastic regime using a longitudinally polarized 5.9 GeV electron beam and a longitudinally polarized proton target at Jefferson Lab with the CEBAF Large Acceptance Spectrometer (CLAS).
The data were collected over a large kinematic phase space and divided into 110 four-dimensional bins of $Q^2$, $x_B$, $-t$ and $\phi$.
Large values of asymmetry moments clearly indicate a substantial contribution to the polarized structure functions from transverse virtual photon amplitudes.
The interpretation of experimental data in terms of generalized parton distributions (GPDs) provides the first insight on the chiral-odd GPDs $\tilde{H}_T$ and $E_T$, and complement previous measurements of unpolarized structure functions sensitive to the GPDs $H_T$ and $\bar E_T$.
These data provides a crucial input for parametrizations of essentially unknown chiral-odd GPDs and will strongly influence existing theoretical calculations based on the handbag formalism.
\end{abstract}

\maketitle

The introduction of generalized parton distributions~(GPDs)~\cite{Mueller:1998fv,Ji:1996nm,Radyushkin:1996nd} defines a new important and far-ranging theoretical framework that allows for the description of the angular momentum components of quarks and gluons in the proton in terms of density distributions in both longitudinal momentum fraction and transverse spatial degrees of freedom.
They provide information on the orbital motion of partons, rendering a three dimensional view of hadron structure \cite{Burkardt:2008jw,Diehl:2003ny}.
Therefore, GPDs are the universal functions that offer an unprecedented opportunity to investigate the nucleon internal structure and provide insight into the hadron at the quark-gluon level.

At leading twist there are eight GPDs~\cite{Diehl:2003ny} for each quark flavor $q$: four correspond to parton helicity conserving (chiral-even) processes, denoted as $H^q$, $E^q$, $\tilde H^q$, $\tilde E^q$, and the remaining four, $H^q_T$, $E^q_T$, $\tilde H^q_T$, $\tilde E^q_T$, correspond to parton helicity-flip (chiral-odd) processes~\cite{PhysRevD.58.054006, Diehl:2001eur}.
The conventional $\bar E_T=2\tilde H_T+E_T$ will be used as well hereafter.
These GPDs can be accessed from the hard exclusive processes such as deeply virtual exclusive photon and meson electroproduction.
%Using the connections between GPDs and helicity amplitudes and applying parity constraints one can single out GPDs from soft matrix elements in hard exclusive processes such as deeply virtual exclusive photon and meson electroproduction.
Deeply virtual pseudoscalar meson electroproduction is sensitive to the chiral-odd GPDs which are less-known than their chiral-even counterparts, because they are not accessible in deeply virtual Compton scattering and, generally, their contributions are suppressed~\cite{Ahmad:2008hp,Goloskokov:2009ia}.
However, their knowledge opens a new avenue to study the partonic structure of the nucleon.
In particular, $H_T$ becomes the quark transversity structure function, $h_1$, in the forward limit,
and it also integrates into the still unknown tensor charge;
the $\bar E_T$ is related to the Boer-Mulders function with its first moment interpreted as the proton's transverse anomalous magnetic moment~\cite{Burkardt2006462, Ahmad:2008hp}.

The unpolarized cross section measurements~\cite{PhysRevLett.109.112001,PhysRevC.90.025205} presented the first evidence that deeply virtual $\pi^0$ electroproduction can be interpreted in terms of the chiral-odd GPDs.
The inclusion of twist-3 components calculated using the chiral-odd GPD parametrizations leads to sizable transverse virtual photon amplitudes and brings theoretical calculations into agreement with experimental data.
However, while the measurements of the unpolarized structure functions and beam spin asymmetries for deeply exclusive $\pi^0$ production have been obtained by the CLAS~\cite{PhysRevLett.109.112001,PhysRevC.90.025205,DeMasi:2007id} and Hall~A Collaborations~\cite{Collaboration:2010kna, PhysRevLett.117.262001}, there are no experimental data available on a longitudinally polarized target.
%They are accessible through the spin asymmetries measurements presented in this paper.
This work presents the first extraction of target and double spin asymmetries for deeply virtual $\pi^0$ production.

The experimental observables in $\pi^0$ electroproduction are connected to the combinations of the different convolutions, defined as~\cite{Goloskokov:2009ia}
\begin{equation}
\langle F\rangle = \int dx \mathcal{H}(x,\xi,Q^2)F(x,\xi,t)
\end{equation}
where $F$ represents generic GPD, $\mathcal H$ is a hard subprocess amplitude, $t$ is a momentum transfer to the nucleon and $\xi$ is a longitudinal momentum fraction transfer.
The unpolarized cross sections contain the combinations of different generalized form factors and, similarly to DVCS, require the polarized structure functions to perform the separation of individual convolutions.
The interpretation of single spin asymmetries, however, is more complicated in comparison with unpolarized cross-sections.
Firstly, the measured asymmetries are the ratios of polarized structure functions and the unpolarized cross section, so the knowledge of unpolarized structure functions is necessary to isolate the polarized contribution.
Secondly, the polarized structure functions are calculated as products of chiral-even and chiral-odd convolutions, complicating the separation of the different contributions.
On the other hand, the double spin asymmetry is well suited for the extraction of chiral-odd GPDs, namely $H_T$, allowing clean separation of $\bar E_T$ and $H_T$ in conjunction with unpolarized target measurements.

Spin asymmetries are defined as a ratio of the difference over the sum of cross sections for opposite helicity configurations and they can be expressed as:

\begin{align}
A_{UL} &= \frac{A_{UL}^{\sin\phi}\sin\phi + A_{UL}^{\sin2\phi}\sin2\phi}{1+A_{UU}^{\cos\phi}\cos\phi + A_{UU}^{\cos2\phi}\cos2\phi}\label{aulphidep},\\
A_{LL} &= \frac{A_{LL}^{const} + A_{LL}^{\cos\phi}\cos\phi}{1+A_{UU}^{\cos\phi}\cos\phi + A_{UU}^{\cos2\phi}\cos2\phi},
\label{allphidep}
\end{align}

\noindent where the first index $U$ ($L$) stands for unpolarized (longitudinally polarized) beam, the second index $U$ ($L$) for the target polarization and $\phi$ is azimuthal angle between the lepton and hadron scattering planes.
$A_{UU}^{\cos\phi}$ and $A_{UU}^{\cos2\phi}$ are connected to the unpolarized structure functions, common for the beam, target and double spin asymmetries,
and $A_{UL}^{\sin\phi}$, $A_{UL}^{\sin2\phi}$, $A_{LL}^{const}$, $A_{LL}^{\cos\phi}$ are connected to the polarized structure functions.

%We present the first measurements of longitudinally polarized target and double spin asymmetries for deeply virtual $\pi^0$ electroproduction over a large phase space.
We present the first measurements of single target and double spin asymmetries for deeply virtual $\pi^0$ electroproduction off the longitudinally polarized protons over a large phase space.
The experiment was carried out in 2009 in Hall~B at Jefferson~Lab, using CLAS~\cite{Mecking:2003zu}, a longitudinally polarized electron beam with average energy of 5.9 GeV and a longitudinally polarized solid ammonia target~\cite{Keith:2003ca}.
The target system, based on a 5~T superconducting magnet and a 1~K $^4$He refrigerating bath, was constructed to polarize protons in paramagnetically doped $^{14}$NH$_3$ along the beam direction via the dynamic nuclear polarization method.
Simultaneously the target's magnetic field serves as an effective shield from M\o ller electrons by focusing them towards the beam line, while allowing detection of photons from $4^\circ$ and maintaining the minimum permitted angle for electrons and protons at 21$^\circ$.
The beam polarization was frequently monitored in M\o ller runs, via the measurement of the asymmetry of elastic electron-electron scattering.
The target polarization was continuously monitored by a Nuclear Magnetic Resonance (NMR) system.
In addition, data were collected using a $^{12}$C target for the purpose of unpolarized nuclear background studies.

The large acceptance of CLAS allowed simultaneous detection of all four final-state particles of the $ep\to ep\pi^0$ and $\pi^0\to\gamma\gamma$ reactions.
The scattered electron was identified by a reconstructed track in the drift chambers and matching it in time with signals in the same CLAS sector of the electromagnetic calorimeter (EC) and the {\v C}erenkov counter.
The cuts on EC energy deposition effectively suppressed the background from negative pions.
The proton was identified as a positively charged particle track in the magnetic field of the superconducting toroidal magnet, passing through the drift chambers with the correct time-of-flight information from the scintillation counters.
The neutral pion decay photons were detected in the EC and the inner calorimeter (IC), which was installed downstream of the target and dedicated to the detection of the photons emitted in the forward direction.
%To eliminate the contamination from M\o ller electrons in the area of the IC close to the beam line the cut $E>1.2-r/6.2$, where $r^2=x^2+y^2$ was applied where $x,y$ are photon's hit coordinates in the IC and $E$ is the energy of the photon.
The photons were detected in the angular range between 4$^\circ$ to 17$^\circ$ in the IC and for angles greater than 21$^\circ$ in the EC.

After the identification of the four particles, the exclusive events from the $ep\to ep\pi^0$ reaction were selected.
With the 4-momenta reconstructed for all final-state particles, the event kinematics is fully known, and energy and momentum conservation can be used to develop the \emph{exclusivity cuts}.
These constraints allow for the rejection of events from unpolarized nuclear background, different channels (e.g. $\eta$, $\rho$ or $\omega$ meson production) and reactions with an additional particle present but undetected.

Three photon-detection topologies exist: (i) both photons detected in the IC, (ii) both photons in the EC and (iii) one photon in the IC and another in the EC.
The experimental resolutions of the kinematic quantities for these topologies were different due to the superior IC resolution, and thus the \emph{exclusivity cuts} were determined independently for each case.
To ensure the exclusivity of $\pi^0$ meson production we used
the 3$\sigma$ cuts extracted from the Gaussian fits of the following four variables:
the missing mass squared $M^2_X(ep)$ of the $(epX)$ system,
the invariant mass of the two photons $M_{\gamma\gamma}$,
the missing energy $E_{ep\gamma\gamma}$ of the $(ep\gamma\gamma)$ system
and the angle $\theta_{\pi^0 X}$ between the measured and the kinematically reconstructed $\pi^0$ meson in the $ep\to epX$ system.

\begin{figure}[t]
  \centering
  \includegraphics[page=1,width=\columnwidth]{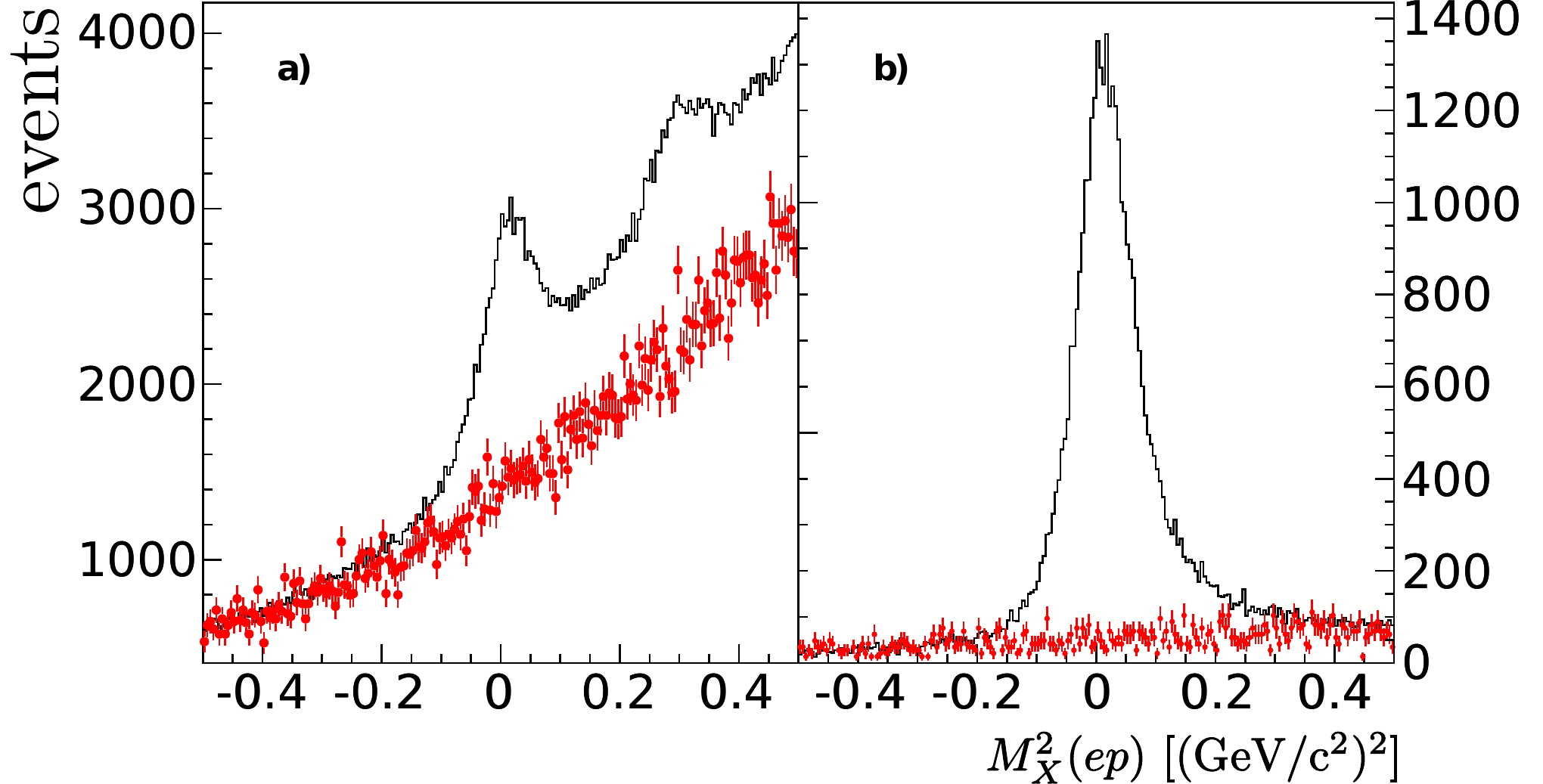}
  \caption{(Color online) Distributions of missing mass squared of the ($ep$) system for the reaction $ep\to ep\pi^0$ before ($a$) and after ($b$) the exclusivity cuts are applied (except $MM^2_X(ep)$ cut itself). The $^{12}$C data (points) are normalized to the $^{14}$NH$_3$ data (line).}
\label{fig:pX2W2}
\end{figure}

Figure~\ref{fig:pX2W2} illustrates the effect of the exclusivity cuts on the missing mass of the $ep$ system in $ep\to epX$.
The contaminations from different meson production and nuclear background are greatly reduced, however even after the application of all exclusivity cuts, the events from nuclear background are still present.
This remaining contamination from $^{14}$N was estimated using the data from carbon runs.
The data from $^{12}$C target were normalized to $^{14}$NH$_3$ by the total charge and corrected for the different areal densities of the target materials.
The variations of the dilution factor with kinematics were too small to parameterize accurately, so constant dilution factors (0.9, 0.94, 0.91) were applied for three topologies (EC-EC, IC-IC and EC-IC).
The contribution from unpolarized nuclear protons was less than 10\% for all topologies.
The small amount of background from accidental photons is visible in Fig.~\ref{fig:pPi02W} under the invariant mass spectrum of the two photons $M_{\gamma\gamma}$ for the three detection topologies.
It was subtracted using the data in the sidebands $(-4.5\sigma,-3\sigma)\cup(3\sigma,4.5\sigma)$ of the $M_{\gamma\gamma}$ distributions independently for each kinematic bin and helicity configuration.
The latter is particularly important because it takes into account any polarization dependent background.

To ensure that the selected events were from the deep-inelastic regime, the kinematic cuts $Q^{2}>1$~(GeV/c)$^{2}$ and $W>2$~GeV/c$^2$ were applied.
$W=\sqrt{(p+q)^2}$ is the $\gamma^{*}p$ invariant mass, where $q$ and $p$ are the four-momenta of the virtual photon and nucleon, and $Q^2=-q^2$.
Then the data were divided into 110 four-dimensional kinematical bins for each of the 4 possible beam/target helicity configurations.
The target and double spin asymmetries were calculated for each kinematic bin as follows:

\begin{figure}[htb]
   \centering
   \includegraphics[page=1,width=0.8\columnwidth]{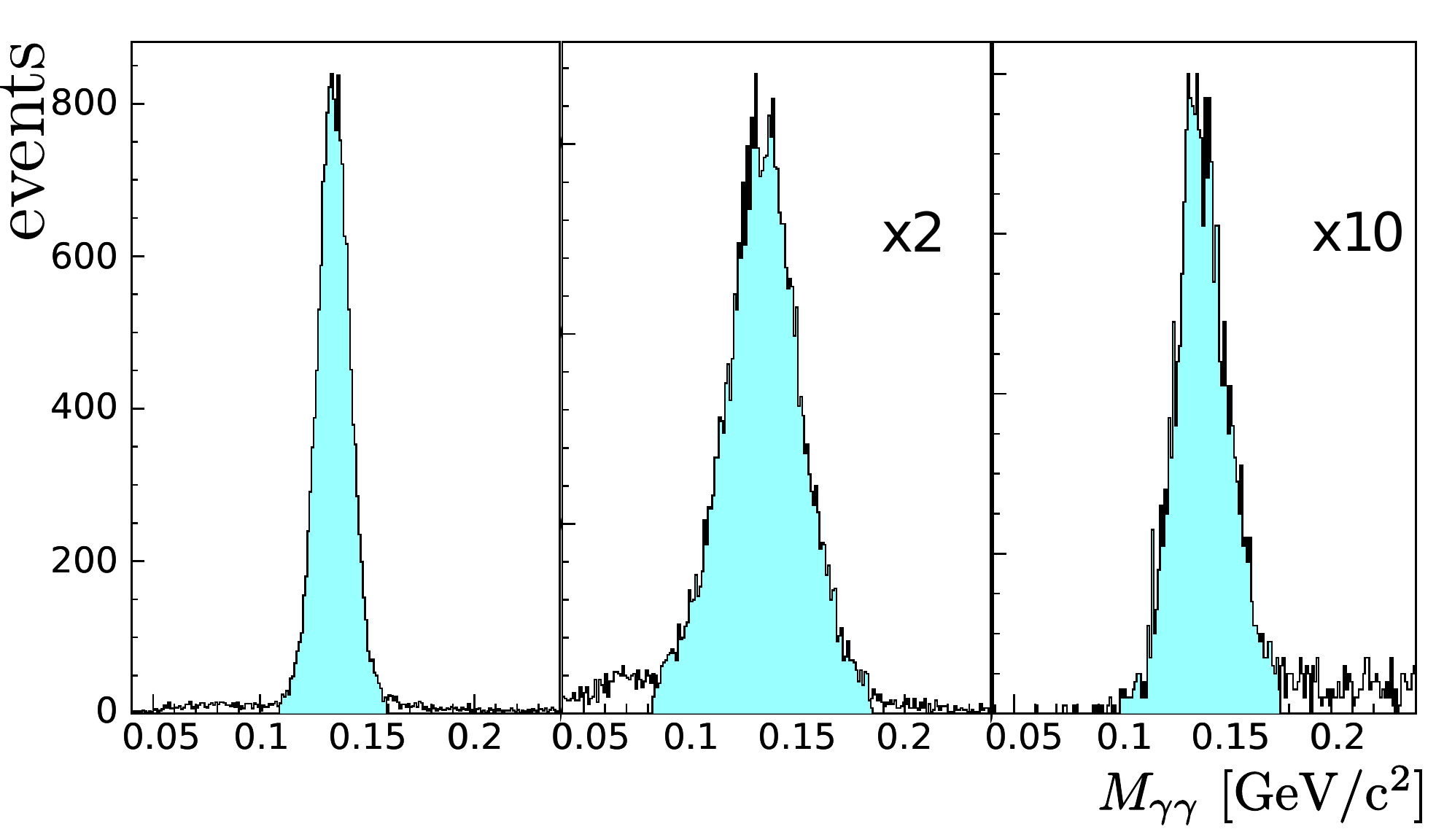}
   \caption{Distributions of invariant mass of the two-photon system for the three different detector configurations: IC-IC, EC-EC and EC-IC from left to right. The cyan areas represent the cuts used for event selection. The last two subfigures show the factors used to scale the number of events in the histograms to the first subfigure.}
\label{fig:pPi02W}
\end{figure}

{\scriptsize
\begin{equation}
A_{UL} = \frac{\sum\limits_i \left(n^{+-}_i + n^{--}_i \right)
-\sum\limits_i \left(n^{++}_i + n^{-+}_i\right)}
{P^-_t\sum\limits_i f_i \left(n^{++}_i + n^{-+}_i \right)
+P^+_t\sum\limits_if_i \left(n^{+-}_i + n^{--}_i\right)},
\end{equation}

\begin{align}
A_{LL} = \frac{1}{P_b}\frac{
\sum\limits_i \left(n^{+-}_i + n^{-+}_i\right)
-\sum\limits_i \left(n^{++}_i + n^{--}_i \right)
}
{P^-_t\sum\limits_i f_i \left(n^{++}_i + n^{-+}_i \right)
+P^+_t\sum\limits_if_i \left(n^{+-}_i + n^{--}_i\right)},
\end{align}
}

\noindent where
$n^{\pm\pm}$ are the numbers of counts for each beam/target helicity configuration, normalized by the corresponding beam charge.
The $i$ index refers to the photon detection topology,
$f_i$ is the corresponding dilution factor,
and $P_t^{\pm}$ are the average values for the positive/negative target polarizations.

The average target polarizations $P^{\pm}_t$ $(P_t^+\simeq80\%,\quad P_t^-\simeq74\%)$ were extracted by dividing the product of beam and target polarizations $P_bP_t$ by the beam polarization $P_b$.
The former was determined by measuring the well-known spin asymmetry in elastic $ep$ scattering~\cite{Donnelly1986247,Perdrisat2007694}.
The latter was measured a few times during the experiment using the M\o ller polarimeter in Hall~B.
The average value was determined to be $84\%\pm2\%$ using the beam polarization measurements weighted by all the events.

The extraction of target and double spin asymmetries for the exclusive $ep\to ep\pi^0$ reaction includes several sources that could induce systematic uncertainties.
The main source was the event selection procedure.
The exclusivity cuts were modified from $2.5\sigma$ to $3.5\sigma$, and the spin asymmetries were re-analyzed for every cut alteration.
The corresponding variations of asymmetries were determined to be 4.4\% on average.
The sideband background subtraction procedure accounted for a systematic uncertainty of 1\%.
To avoid systematic uncertainty associated with NMR measurements the product of beam and target polarizations was extracted from the exclusive $ep$ elastic scattering.
Therefore, combined with the dilution factor, the uncertainties of the beam and the target polarizations lead to an overall normalization uncertainty of 3\% for double spin asymmetry and 5\% for target spin asymmetry.
The acceptance and binning effects were studied through careful Monte-Carlo simulation, and both effects were found to be negligible.
The individual uncertainties for each kinematic bin were added in quadrature, and their values, with an average of 4.5\%, were found to be smaller than statistical uncertainties for the most of bins.

\begin{figure*}[t]
\includegraphics[width=\textwidth]{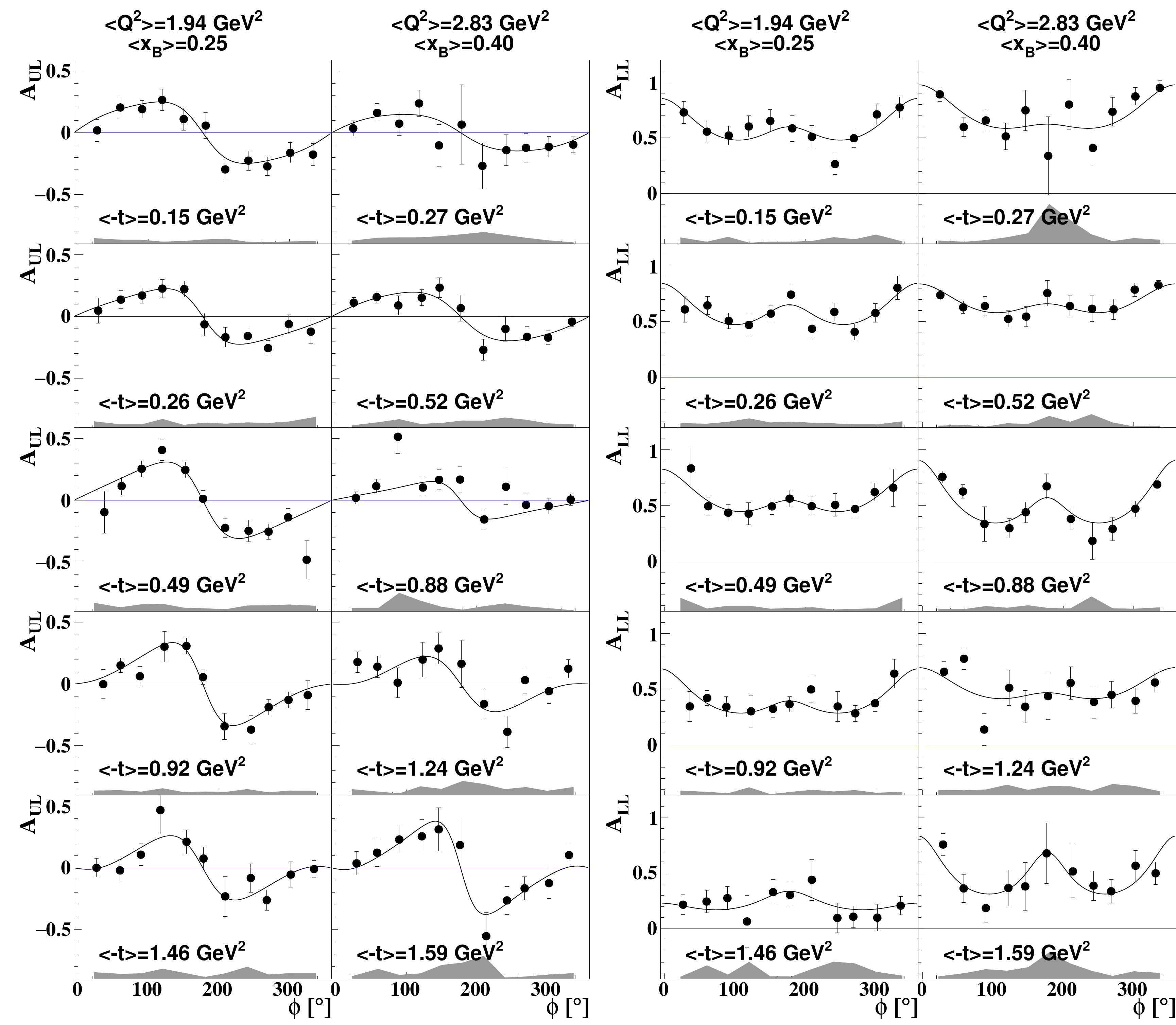}
 \caption{
  Target and double spin asymmetries for deep exclusive $\pi^0$ production plotted as a function of $\phi$ for each kinematic bin in $\left(Q^2,x_B\right)$ space and $-t$ range.
  The curves are simultaneous fit results described in the text.
  The shaded bands represent the overall systematic uncertainties.
  The latter exhibits strong variation with $\phi$ in certain kinematic bins due to unseparable statistical fluctuations inflating the estimate of the systematic errors.
  The bins with low statistics are particularly affected by them.
 }\label{fig:phidep}
\end{figure*}

The target and double spin asymmetries for exclusive $\pi^0$ production were measured over a wide kinematic range with $1<Q^2<5$~(GeV/c)$^2$, $0.1<x_B<0.6$, and $0<-t<2$~(GeV/c)$^2$,
where $x_B=\frac{Q^2}{2pq}$ is the Bjorken variable, $t=(p-p^\prime)^2$ is the momentum transfer to the nucleon, and $p$ and $p^\prime$ are the initial and final four-momenta of the nucleon.
The data were divided into two bins in the $(Q^2,x_B)$ space, five bins in $-t$ and eleven $\phi$ bins with the measured asymmetries shown as a function of $\phi$ in Fig.~\ref{fig:phidep}.
The measurements exhibit strong azimuthal dependence for the target spin asymmetries with significant amplitudes of the $\sin\phi$ moments and a large constant term for the double spin asymmetries.

\begin{figure}[tb]
\centering
\includegraphics[width=0.95\columnwidth]{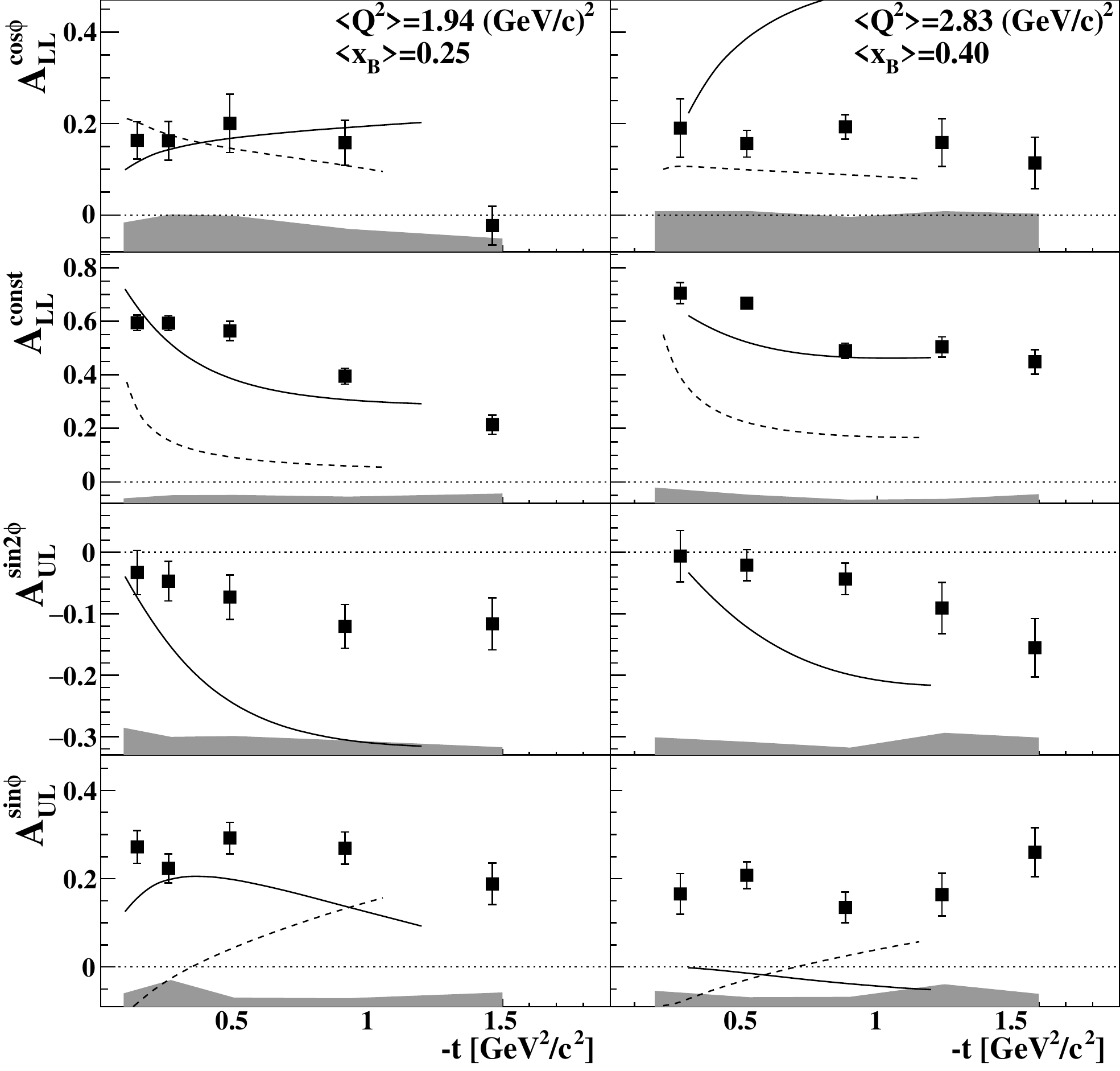}
\caption{\label{fig:moments}
The target and double spin asymmetry moments of exclusive $\pi^{0}$ electroproduction as a function of $-t$ for 2 bins in the ($Q^{2}, x_B$) plane (left and right columns).
The shaded bands represent the systematic uncertainties (including fit uncertainties), and the curves show the predictions from two GPD formalisms:
GK~\cite{Goloskokov:2011rd} (dashed) and GGL~\cite{PhysRevD.91.114013} (solid).
Note that the $A_{UL}^{\sin2\phi}$ moment is zero in the GK model, and therefore is not shown.
}
\end{figure}

The measured beam, target and double spin asymmetries were fitted simultaneously using six free parameters: $A_{LU}^{\sin\phi}$, $A_{UL}^{\sin\phi}$, $A_{UL}^{\sin2\phi}$, $A_{LL}^{const}$, $A_{LL}^{\cos\phi}$ and $A_{UU}^{\cos2\phi}$ according to the Eqs.~\ref{aulphidep}~and~\ref{allphidep} to describe their azimuthal dependence.
Simultaneous fit exploits the fact that the three asymmetries have the same denominator and constrains the common terms to be the same for the three different observables.
The beam spin asymmetry was extracted in addition to the target and double spin asymmetries.
This observable is important in the simultaneous fit to better constrain the unpolarized term $A_{UU}^{\cos2\phi}$ in the denominator common for the different polarized observables.
Both $A_{LU}^{\sin\phi}$ and $A_{UU}^{\cos2\phi}$ are by-products of the measurement and much better constrained from previous experiments with an unpolarized hydrogen target.
Due to the limited statistics, the term $A_{UU}^{\cos\phi}=\sqrt{2\epsilon(1+\epsilon)}\sigma_{LT}/\sigma_{0}$ was fixed using the structure functions reported by CLAS in~\cite{PhysRevLett.109.112001}.
The correlations of numerator terms with $A_{UU}^{\cos\phi}$ and $A_{UU}^{\cos2\phi}$ were studied and found to be small for all asymmetry moments except $A_{LL}^{\cos\phi}$ term.
To verify the stability of moments extraction the fit was performed with and without $A_{UU}^{\cos\phi}$ term.
Only $A_{LL}^{\cos\phi}$ exhibits positive correlation with unpolarized terms as evident from large systematic uncertainties on Fig~\ref{fig:moments} driven by uncertainties on the denominator parameters.

In Fig.~\ref{fig:moments} the measured asymmetry moments for $\pi^0$ electroproduction are plotted as a function of $-t$ in each ($Q^{2}, x_B$) bin, where each kinematic value is calculated as event weighted average.
The theoretical predictions from two GPD-based approaches, GK~\cite{Goloskokov:2011rd} and GGL~\cite{PhysRevD.91.114013}, are also included.
They both calculate the contributions from the transverse virtual photon amplitudes using chiral-odd GPDs with $-t$ dependence, incorporated from Regge phenomenology, but differ in the GPD parametrization methods.
GGL provides the chiral-odd GPD parametrization via linear relations to chiral-even GPDs under parity and charge conjugation symmetries in a Regge-ized diquark model.
This approach allows the model to overcome the issue that very few constraints on chiral-odd GPDs exist, while chiral-even GPDs can be relatively well-constrained using deep inelastic scattering, nucleon form-factor and DVCS measurements.
In the GK model, chiral-odd GPDs are constructed from the double distributions and constrained using the latest results from lattice QCD~\cite{PhysRevLett.98.222001} and transversity parton distribution functions~\cite{Anselmino200998} with the emphasis on $H_T$ and $\bar E_T$, while the contribution from the other chiral-odd GPDs are considered negligible.

Within the approximation of GK model~\cite{Goloskokov:2011rd} the $A^{const}_{LL}$ component is defined as:
\begin{equation}
A^{const}_{LL}\sigma_0=\sqrt{1-\epsilon^2}\frac{4\pi\alpha}{k}\frac{\mu_\pi^2}{Q^4}(1-\xi^2)\left|\langle H_T\rangle\right|^2
\end{equation}
A rather straightforward interpretation is obtained for $A_{LL}^{const}$ component, which contains contributions from the chiral-odd GPDs only.
It is expected to be determined by $\langle H_T\rangle$ with a negligible contribution from $\langle\tilde H_T\rangle$.
The data do not decrease near threshold which is indicative of $\langle H_T\rangle$ dominance, especially at low $-t$ region,
and both models display a rise of $A_{LL}^{const}$ in the same region since they rely on $\langle H_T\rangle$ to describe it.
Furthermore, the extraction of $A_{LL}^{const}$ term is very stable due to the absence of $\phi$ dependence and, therefore, provides a reliable experimental observable to constrain GPD $H_T$.
Additionally, and perhaps more importantly, the double spin asymmetry measurements provide an independent test of the existing GPDs models probing the underlying assumption of $H_T$ and $\bar E_T$ dominance.
The first approach to combine $A_{LL}^{const}$ term with $\sigma_0$ and $\sigma_{TT}$ previously measured by CLAS was conducted in ~\cite{Kroll:1602.03803}.
The constant term was calculated within GK model framework using the $H_T$ and $\bar E_T$ convolutions extracted from unpolarized cross section measurements~\cite{PhysRevLett.109.112001,PhysRevC.90.025205} and compared to the measurements presented in this work.
The agreement within error bars demonstrates the major contribution from $\langle H_T\rangle$.

The large magnitudes of $A^{\cos\phi}_{LL}$ and $A^{\sin\phi}_{UL}$ components suggest sizable contributions from the chiral-odd GPDs through the interference of transverse and longitudinal virtual photons amplitudes.
The calculations of these contributions are complicated, largely due to the unknown phases between interfering terms~\cite{Goloskokov:2011rd,PhysRevD.91.114013}.
The $A^{\sin\phi}_{UL}$ term exhibits a relatively flat $-t$ dependence similar to the observed dependence of the beam-spin asymmetry~\cite{DeMasi:2007id}, but with a factor of three larger magnitude.
Note that both terms are dominated by $\langle\tilde H\rangle^*\langle E_T\rangle$, but the target spin asymmetry is also enhanced by $\langle\tilde H\rangle^*\langle\tilde H_T\rangle$ and the beam spin asymmetry is reduced by a kinematic factor.
The $A_{UL}^{\sin2\phi}$ component is determined by chiral-odd GPDs, namely $E_T$ and $\tilde H_T$, providing the means to disentangle these two GPDs and improve the parameterization of their combination $\bar E_T$~\cite{PhysRevD.91.114013}.

In conclusion, for the first time target and double spin asymmetries from deeply virtual $\pi^0$ meson production were extracted over a wide range of $Q^2$, $x_B$ and $-t$.
The measurements shown in Fig.~\ref{fig:moments} are significantly different from zero in all kinematic bins.
Our data provide a set of new observables in the kinematic range of $t/Q^2$ where higher twist contributions may be significant.
The overall comparison of experimental measurements with different theoretical calculations using the leading order demonstrates the importance of our results to improve parameterization of the GPD $H_T$.
They indicate strong sensitivity to the practically unknown $-t$ dependencies of the underlying chiral-odd GPDs, which may shed light on the role of higher twist contributions.
Combined with the unpolarized structure function measurements and beam spin asymmetry results for $\pi^0$ production from CLAS~\cite{DeMasi:2007id,PhysRevLett.109.112001,PhysRevC.90.025205},
these data provide important constraints for the parameterizations of GPDs $H_T$ and $\bar E_T$, giving the first insight to the transverse space distributions of transversely polarized quarks~\cite{Diehl:2005jf,Hannafious:2008dx}.

%%%%%%%%%%%%%%%%%%%%%%%%%%%%%%%%%%%%%%%%%%%%%%%%%%%%%%%%%%%%%%%%%%%%%%%%%%%%%%%%%%%%%%%%
%%%%%%%%%%%%%%%%%%%%%%%%%%%%%%%%%%%%%%%%%%%%%%%%%%%%%%%%%%%%%%%%%%%%%%%%%%%%%%%%%%%%%%%%
%%%%%%%%%%%%%%%%%%%%%%%%%%%%%%%%%%%%%%%%%%%%%%%%%%%%%%%%%%%%%%%%%%%%%%%%%%%%%%%%%%%%%%%%
\section*{Acknowledgments}
We acknowledge the outstanding efforts of the staff of the Accelerator and Physics Divisions at JLab.
This work was supported in part by the U.S. Department of Energy and National Science Foundation, the French Centre National de la Recherche Scientifique and Commissariat \`{a} l'Energie Atomique, the Italian Istituto Nazionale di Fisica Nucleare, the National Research Foundation of Korea and the U.K. Engineering and Physical Science Research Council.
% Emmy Noether grant from the Deutsche Forschungs gemeinschaft
Jefferson Science Associates (JSA) operates the Thomas Jefferson National Accelerator Facility for the United States Department of Energy under contract DE-AC05-06OR23177.

%\bibliographystyle{elsarticle-num-names}
%\bibliography{biblio}

\end{document}